\documentstyle[sprocl,11pt,epsfig,twoside]{article}	
\newcommand{\ifm}[1]{\relax\ifmmode #1\else $#1$\fi}
 
\newcommand{\beq   }{\begin{equation}}
\newcommand{\eeq   }{\end{equation}}
\newcommand{\beqn  }{\begin{eqnarray}}
\newcommand{\eeqn  }{\end{eqnarray}}
\newcommand{\bi    }{\begin{itemize}}
\newcommand{\ei    }{\end{itemize}}
\newcommand{\bc    }{\begin{center}}
\newcommand{\ec    }{\end{center}}
\newcommand{\bd    }{\begin{description}}
\newcommand{\ed    }{\end{description}}
\newcommand{\bHuge }{\begin{Huge}}
\newcommand{\bhuge }{\begin{huge}}
\newcommand{\bLARGE}{\begin{LARGE}}
\newcommand{\bLarge}{\begin{Large}}
\newcommand{\blarge}{\begin{large}}
\newcommand{\eHuge }{\end{Huge}}
\newcommand{\ehuge }{\end{huge}}
\newcommand{\eLARGE}{\end{LARGE}}
\newcommand{\eLarge}{\end{Large}}
\newcommand{\elarge}{\end{large}}

\def \gtsim    {\relax\ifmmode{\mathrel{\mathpalette\oversim >}}
                  \else{$\mathrel{\mathpalette\oversim >}$}\fi}
\def \ltsim    {\relax\ifmmode{\mathrel{\mathpalette\oversim <}}
                  \else{$\mathrel{\mathpalette\oversim <}$}\fi}
\def\oversim#1#2{\lower4pt\vbox{\baselineskip0pt \lineskip1.5pt
            \ialign{$\mathsurround=0pt#1\hfil##\hfil$\crcr#2\crcr\sim\crcr}}}
\newcommand{\gev}  {\mbox{${\rm GeV}$}}

\newcommand{\pgev} {\mbox{${\rm GeV}/c$}}

\newcommand{\mgev} {\mbox{${\rm GeV}/c^2$}}

\newcommand{\invfb}{\mbox{${\rm fb}^{-1}$}}

\newcommand{\degr  }{\mbox{$^{\circ}$}}
\newcommand{\pt}  {\mbox{$p_{T}$}}

\newcommand{\mpt} {\mbox{${p\!\!\!/_T}$}}
\newcommand{\azero}{\ifm{A_0}}
\newcommand{\tanb}{\ifm{\tan\beta}}
\newcommand{\mzero}{\ifm{m_0}}
\newcommand{\mhalf}{\ifm{m_{1/2}}}
\newcommand{ \stauone}  {\mbox{$\tilde{\tau}_{1}$}}
\newcommand{ \stauonep} {\mbox{$\tilde{\tau}_{1}^{+}$}}
\newcommand{ \stauonem} {\mbox{$\tilde{\tau}_{1}^{-}$}}

\newcommand{ \schionezero }{\mbox{$\tilde{\chi}_{1}^{0}$}}
\newcommand{ \schitwozero }{\mbox{$\tilde{\chi}_{2}^{0}$}}

\newcommand{ \schionepm }{\mbox{$\tilde{\chi}_{1}^{\pm}$}}

\newcommand{ \bsmumu }{\mbox{$B_{s} \to \mu^+ \mu^-$}}

\def \PRL      {Phys. Rev. Lett.~}

\def \PRD      {Phys. Rev. D}

\def \PLB      {Phys. Lett. B}
\def \NPB      {Nucl. Phys. B}

\def \etal     {\relax\ifmmode{et \; al.}\else{$et \; al.$}\fi}

\def \calR     {\relax\ifmmode{{\cal R}}\else{${\cal R}$}\fi}
\def \Dzero    {\relax\ifmmode{{\rm D\O}}\else{D\O}\fi}
\def \DzeroC   {\relax\ifmmode{{\rm D\O\ Collaboration}}
	\else{D\O\ Collaboration}\fi}
\def \CDFC   {\relax\ifmmode{{\rm CDF Collaboration}}
	\else{CDF Collaboration}\fi}
\def \CDFIIC   {\relax\ifmmode{{\rm CDF II Collaboration}}
	\else{CDF II Collaboration}\fi}
\begin{document}
\setcounter{page}{1}
\title{Probing mSUGRA Models at Linear 
Colliders\footnote{T.K. and V.K. are  supported
	by DOE grant  DE-FG03-95ER40917, and 
	R.A. and B.D. 
 	by NSF grant PHY-0101015.}} 
\author{T. Kamon,\footnote{e-mail address: t-kamon@tamu.edu}
	R. Arnowitt, B. Dutta,\footnote{Present Address:
	{\it Department of Physics, 
	University of Regina, Regina SK, S4S 0A2, Canada} }
	  and V. Khotilovich
\\
        {\it Department of Physics, Texas A\&M University, 
	College Station, TX 77843-4242, USA}
}
\maketitle
\begin{abstract}
A  feasibility study
of a 500-GeV linear collider is presented for
mSUGRA models in co-annihilation region.
We find an active mask is critical
to suppress 
$e^+ e^- \rightarrow e^+ e^- \tau^+ \tau^-$ events
to probe the models.
\end{abstract}

\section{Feasibility of 500-GeV Linear Collider}
\begin{figure}
\begin{center}
\epsfig{file=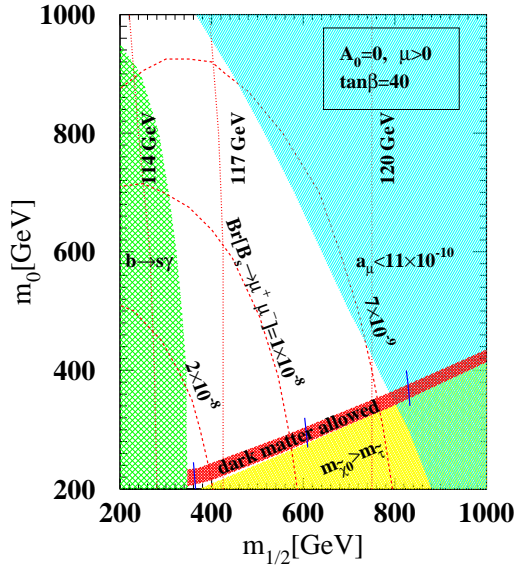,height=3.in,angle=0}
\caption{Parameter space, in $m_0$-$m_{1/2}$ plane of mSUGRA models
($\tanb$ = 40),
allowed by relic density of $\schionezero$'s
as cold dark matter.\protect\cite{Tevatron_Bsmumu_Prospects}
Also shown are branching ratios for \bsmumu, where
three dashed lines indicate
$Br$ = $2\times 10^{-8}$, $1\times 10^{-8}$, and $7\times 10^{-9}$.
The Tevatron Run II with 15 \invfb\
will be accessible to $1\times 10^{-8}$.
The vertical dotted lines label Higgs masses.
The three short solid lines indicate the $\sigma_{p \tilde\chi^0_1}$ values 
(from left:  0.03 $\times 10^{-6}$ pb,  
0.002 $\times 10^{-6}$ pb, 0.001 $\times 10^{-6}$ pb).  
}
\label{fig:LC_mSUGRA_tanB40}
\end{center}
\end{figure}

The grand unification of the three gauge
coupling constants arise naturally
in minimal SUGRA (mSUGRA) models 
with universal soft breaking parameters
($\mhalf$, $\mzero$, $\azero$, $\tan\beta$,
sign of $\mu$).\cite{Chamseddine:jx,sugra2}
The experimental data to provide 
the most significant constraints
are 
the Higgs mass bound, 
the $b \rightarrow s \gamma$ branching ratio, 
and (possibly) the muon magnetic moment anomaly ($a_\mu$)
along with
the amount of relic density of the lightest neutralino
($\schionezero$) as cold dark matter.
These constraints produce a lower bound of $\mhalf\ \gtsim\ 300\ \gev$
(Higgs mass and $b \to s \gamma$)
and an upper bound of $\mhalf\ \ltsim\ 900\ \gev$ ($a_\mu$),
making the SUSY spectrum well accessible 
to the LHC.\cite{mSUGRA_cdm}\
The relic density constraints limit
the most part of the allowed parameter space
to narrow band where $\stauone$-$\schionezero$
co-annihilation occurs.\cite{coannihilation1,coannihilation2}\
See, for example, 
Fig.~\ref{fig:LC_mSUGRA_tanB40}.\cite{Tevatron_Bsmumu_Prospects}\
Further, at large $\tanb$ (e.g., 40),
$M_{\stauone}$ becomes
the next to lightest SUSY particle and
all other SUSY particles would be too heavy to produce in pair
at $e^+ e^-$ linear colliders (LCs).
The only accessible production at 500-GeV LCs could be
$\schitwozero\schionezero$,
$\stauonep\stauonem$ and $\schionezero\schionezero$,
where
$M_{\schionepm} > 250\ \gev$ and
$\Delta M \equiv M_{\stauone} - M_{\schionezero}$ = 5 - 30 \gev.
Furthermore, the $\schitwozero \to \stauone \tau$ decay
is dominant.
This guides us to the conclusion
that one of the most critical experimental sigantures is
$\tau^+ \tau^-$ plus missing energy.
We present a feasibility study of a 500-GeV LC for the signature
of two hadronically-decaying $\tau$ lepton ($\tau_h$)
plus missing energy.

We generate (i)~$\schitwozero \schionezero$
and $\stauonep \stauonem$ events via ISAJET and
(ii)~SM four-fermion final states of
$\nu_{e,\mu,\tau} \bar{\nu}_{e,\mu,\tau} \tau^+ \tau^-$
from weak process
and $e^+ e^- \tau^+\tau^-$ ($\pt_{\tau} > 3\ \pgev$)
from two-photon ($\gamma\gamma$) process
via WPHACT.\cite{MCtool}\
For the SUSY events, we choose three mSUGRA points,
as listed in Table~\ref{tab:mSUGRApoints}.
Those events are simulated and analyzed with
the LCDRoot software package.\cite{NLCsim}\
The electron-beam
polarization is set to be ${\cal P}(e^-)$ = $0.9$ (L.H)
to enchance the $\schitwozero\schionezero$ signal
and the opposite polarization ${\cal P}(e^-)$ = $-0.9$ (R.H) 
is used for the $\stauonep\stauonem$ signal extraction.
See Table~\ref{tab:RH_vs_LH}.

\begin{table}
\caption{Three mSUGRA points 
for $\mu > 0, \tanb = 40$, and $\azero = 0$
 by ISAJET.\protect\cite{MCtool} }
\begin{center}
\begin{tabular}{l cc cccc }
\hline
\hline
  & $\mhalf$  & $\mzero$  &
\multicolumn{1}{c}{$M_{\schitwozero}$}  &
\multicolumn{1}{c}{$M_{\stauone}$}   &
\multicolumn{1}{c}{$M_{\schionezero}$}     &
\multicolumn{1}{c}{$\Delta M$}   \\
 & & & & & &
\multicolumn{1}{c}{$( \equiv M_{\stauone} - M_{\schionezero} )$}   \\
\hline
A1 & 360 & 210 & 266 & 149.9 & 144.2 &  5.7\\
A2 & 360 & 215 & 266 & 154.8 & 144.2 & 10.6 \\
A3 & 360 & 225 & 266 & 164.3 & 144.4 & 20.2 \\
\hline
\hline
\end{tabular}
\end{center}
\label{tab:mSUGRApoints}
\end{table}

\begin{table}
\caption{Production cross sections (fb) for the SM (excluding
$\gamma\gamma$ process) and
SUSY events in $e^+ e^-$ collisions with different
electron-beam polarizations 
(${\cal P}(e^-)$ = $-0.9$, 0, and 0.9)
at $\sqrt{s}$ = 500 GeV.
}
\begin{center}
\begin{tabular}{l cc cc cc }
\hline
\hline
 ${\cal P}(e^-)$  & 
\multicolumn{2}{c}{$-0.9$ (R.H.)} & 
\multicolumn{2}{c}{0} & 
\multicolumn{2}{c}{$+0.9$ (L.H.)} \\
\hline
SM (excluding $\gamma\gamma$) & 
\multicolumn{2}{c}{7.84} & 
\multicolumn{2}{c}{48.9} & 
\multicolumn{2}{c}{89.8} \\
\hline
SUSY & 
$\schionezero \schitwozero$ & $\stauone \stauone$ &
$\schionezero \schitwozero$ & $\stauone \stauone$ &
$\schionezero \schitwozero$ & $\stauone \stauone$ \\
\multicolumn{1}{r}{A1} & 0.529 & 26.4 & 3.39 & 19.6 & 7.10 & 12.8 \\
\multicolumn{1}{r}{A2} & 0.520 & 24.4 & 3.31 & 18.4 & 6.91 & 11.8 \\
\multicolumn{1}{r}{A3} & 0.501 & 21.1 & 3.15 & 15.8 & 6.62 & 10.3 \\
\hline
\hline
\end{tabular}
\end{center}
\label{tab:RH_vs_LH}
\end{table}

For the L.H. case, we optimize the selection cuts
for $\schitwozero\schionezero$ events
over the $\stauonep\stauonem$ and SM events:
(1)~$N_{jet} = 2$ (JADE Y-cut $\ge$ 0.0025), where each jet passes
a $\tau_h$ identification (1 or 3 tracks; $M_{jet} < 2\ \mgev$)
with $E_{jet} > 3\ \gev$;
(2)~Opposite charge between two jets ($q_{j_1} \cdot q_{j_2} = -1$);
(3)~$-q_{j} \cdot \cos \theta_{j} < 0.75$;
(4)~Missing \pt\ ($\equiv \mpt$) $> 5, 10$, or $20\ \pgev$;
(5)~$\cos \theta(\vec{p}_{j_2}, \vec{p}_{vis}) > -0.6$;
(6)~Accoplanarity $> 40\degr$;
(7)~Vetoing event if there is any EM cluster with $E_{cluster} > 2\ \gev$
	in $0.995 < | \cos\theta_{cluster}| < 0.9$
	($5.8\degr < \theta_{cluster} < 25.8\degr$),
or  any cluster with $E_{cluster} > 100\ \gev$ 
in $2(1)\degr < \theta_{cluster} < 5.8\degr$, 
or electron/positron with $E_e > 1.5\ \gev$ in $|\cos\theta_e| < 0.9$.
In the present study, the same cuts are used for the R.H. case.
The results are summarized in
Table~\ref{tab:LC500_Results}.
With an active mask of covering down to 2\degr,
the LC detector would be able to test
mSUGRA models with small $\Delta M$, and being better
with 1\degr\ mask.\\[-.25in]

\begin{table}
\caption{Number of events expected at 500-GeV LC
with 500 \invfb.
Three mSUGRA points
are examined.
Boldfaced numbers for L.H. (R.H.) electron-beams
indicate $>5\sigma$-discovery sensitivities for
$\schitwozero\schionezero$ ($\stauonep\stauonem$)
over $\stauonep\stauonem$+SM ($\schitwozero\schionezero$+SM) background events. 
The SM ``weak'' process 
includes all $\nu_{e,\mu,\tau} \bar{\nu}_{e,\mu,\tau} \tau^+ \tau^-$
final states, while the SM $\gamma\gamma$
is the $ee\tau\tau$ ($\pt_{\tau} >3\ \pgev$) final state.
}
\begin{center}
\begin{tabular}{l lr | rrr | rrr}
\hline
\hline
  & & & 	\multicolumn{3}{c|}{${\cal P}(e^-)=0.9$(L.H.)} &  
	\multicolumn{3}{c}{${\cal P}(e^-)=-0.9$(R.H.)} \\
Process &   &  & $\mpt^{min}$ = 5  & 10 & 20 & 5 & 10 & 20 \\
\hline
$\schitwozero\schionezero$ 
	& A1 & & {\bf 609} & {\bf 545} & {\bf 402} &
			48 & 42 & 31 \\
	& A2 & & {\bf 952} & {\bf 873} & {\bf 626} &
			71 & 65 & 47 \\
	& A3 & & {\bf 1180} & {\bf 1113} & {\bf 835} &
			90 & 85 & 63 \\
\hline
$\stauonep\stauonem$ 
	& A1 & & 170 & 16 & 0 & 
			{\bf 353} & 37 & 0\\
	& A2 & & 687 & 394 & 30 & 
			{\bf 1442} & {\bf 817} & 59 \\
	& A3 & & 1198 & 994 & 438 & 
			{\bf 2510} & {\bf 2098} & {\bf 939} \\
\hline
SM weak    &  &    & 2486 & 2320 & 1797 & 391 & 373 & 312 \\
SM $\gamma\gamma$ & ~2-5.8\degr\ mask & & 1507 & 13 & 0 & 1507 & 13 & 0\\
             & [1-5.8\degr\ mask] &  & [38] & [2] & [0] & [38] & [2] & [0]\\
             & [no mask] &  & [11400] & [3020] & [89] & [11400] & [3020] & [89]\\
\hline
\hline
\end{tabular}
\end{center}
\label{tab:LC500_Results}
\end{table}

\section{Conclusion}

We have examined a feasibility of a 500-GeV LC 
in probing three mSUGRA points in co-annihilation region.
Our preliminary result showed that
an active mask is very critical for the study,
to suppress
$e^+ e^- \rightarrow e^+ e^- \tau^+ \tau^-$ events.
In future study, we will include
$e^+ e^- \rightarrow e^+ e^- q \bar{q}$
where the quark jets are misidentified as $\tau_h$'s.


\begin{thebibliography}{99}
\bibitem{NLCsim}
LCDRoot Package v3.5 with
LD Mar01 detector parametrization
of EM and HAD calorimeters ($|\cos\theta | < 0.995$)
and tracking system ($|\cos\theta | < 0.9$).
\end{thebibliography}
\end{document}